\documentclass[12pt]{article} % single column, double spaced

\usepackage{ol2}
\usepackage[draft]{hyperref}
\usepackage{amsmath}
\begin{document}

%\twocolumn[ %% activate for two-column option

\title{Principal modes in fiber amplifiers}

%% For REVTeX it is possible to automate superscript and e-mail callouts with the superscriptaddress option; see REVTeX4 documentation.

\author{Moti Fridman$^{1*}$, Micha Nixon$^1$, Mark Dubinskii$^2$, Asher A. Friesem$^1$, Nir Davidson$^1$}

\address{
$^1$ Dept. of Physics of Complex Systems, Weizmann Institute of Science, Rehovot 76100, Israel \\
$^2$ U.S. Army Research Laboratory, Adelphi, Maryland 20783, USA  \\
$^*$ Corresponding author: moti.fridman@weizmann.ac.il

}

\begin{abstract}
The dynamics of the state of polarization in single mode and multimode fiber amplifiers are presented. The experimental results reveal that although the state of polarizations at the output can vary over a large range when changing the temperatures of the fiber amplifiers, the variations are significantly reduced when resorting to the principal states of polarization in single mode fiber amplifiers and principal modes in multimode fiber amplifiers.
\end{abstract}


% ] %% activate for two-column option

\noindent

It is well known that in general optical fibers have complicated polarization characteristics, like random birefringence and polarization mode dispersion which limit the performance of optical communication systems~\cite{PSP86, Kogelnik}. To overcome such limitations, investigators have resorted to using polarization maintaining fibers~\cite{PMmultimode1, PMmultimode2, PMmultimode3} or operating at the principal state of polarization~\cite{dynamicPSP, dynamicPSP2, dynamicPSP3}. The situation becomes more complicated when dealing with non-polarization maintaining multimode fibers, which were theoretically investigated to show that principal modes are best for obtaining least dispersion~\cite{KahnOL, Kahn2, Kahn3}.

In this letter, we present novel experimental results on the dynamics of the state of polarization (SOP) in single mode and multimode fiber amplifiers as a function of temperature. We show that the output SOP strongly depends on the pump power or equivalently the temperature, but such dependence can be significantly alleviated by resorting to the principal state of polarizations (PSP) in single mode fiber amplifiers and principal modes (PM) in multimode fiber amplifiers. This is especially important for applications where the repetition rate of the laser changes, since changing the repetition rates also changes the temperature of the fiber amplifier. Such applications are marking, engraving, micro-machining, printing, material processing, and many others~\cite{app1, app2}. In these applications the PSP or the PM vary slowly and can be tracked while the temperature of the fiber amplifiers fluctuates rapidly.

%Fiber lasers and fiber amplifiers with their high efficiencies, robustness, compactness, and low costs draw much attention over the last years. Much effort was investigated in developing PM fiber amplifiers, so the polarization at the output will be stable as a function of the input wavelength. However, these fibers are expensive, complicated to manufacture and need to be spliced very accurately. It is known that in non-PM single mode fibers there are two states of polarizations, namely principal state of polarizations(PSP), which are not sensitive to variation in the input wavelength. In multimode fibers these states are called principal modes~(PM)~\cite{KahnOL, Kahn2}. Here we show experimentally that variations in the pump power of the amplifiers induce thermal variations which are equivalent to changes in the input wavelength. Therefore, operating fiber amplifiers in the PSP for single mode fiber amplifiers or at the PM for multimode fiber amplifiers, makes the output polarization and the output mode invariant for variations in the pump power, in the temperature of the fiber amplifier and in the input wavelength.

\begin{figure}[htb]
\centerline{\includegraphics[width=8.3cm]{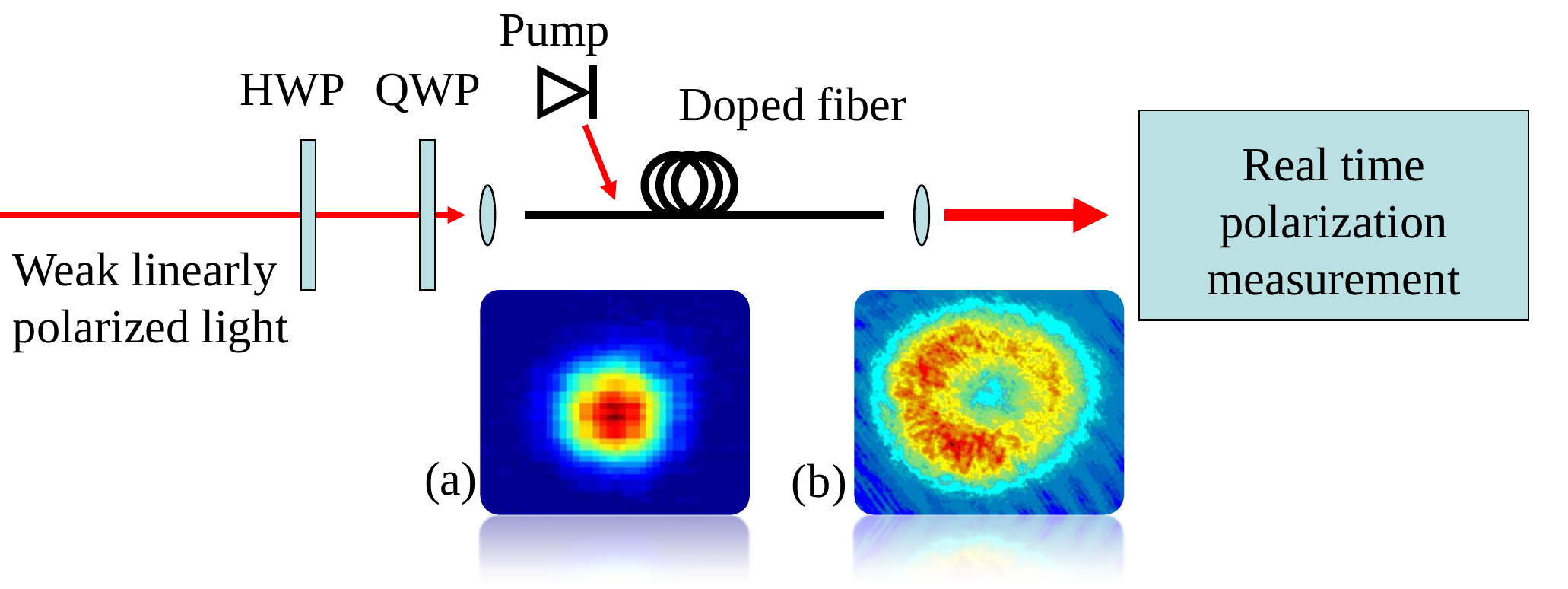}}
\caption{\label{system} Experimental configuration for measuring the SOP at the output of a fiber amplifier. HWP - half wave plate; QWP - quarter wave plate. Inset (a) shows the Gaussian shape of the amplified spontaneous emission from a single mode fiber amplifier; inset (b) shows the doughnut shape from the multimode fiber amplifier.}
\end{figure}

The experimental configuration for measuring the SOP at the output of a fiber amplifier as a function of the pump power is schematically shown in Fig.~\ref{system}. A near Gaussian linearly polarized light at wavelength of $1064nm$ propagates through a half wave plate and a quarter wave plate in order to obtain any desired light polarization at the input. The light is then injected into a $10m$ long Ytterbium doped fiber amplifier that is pumped from the side with a $915nm$ diode laser of about $2W$. The output power is less than $500mW$ so nonlinear effects can be neglected. We also assume that the losses are negligible and the birefringence is constant across the relatively small fiber core. The full SOP at the output as a function of space and time is measured using a real-time space-variant polarization measurement system~\cite{RTSVPM}. Also shown the amplified spontaneous emission from a single mode fiber amplifier [inset(a)] and from a multimode fiber amplifier [inset (b)].

\begin{figure}[htb]
\centerline{\includegraphics[width=8.3cm]{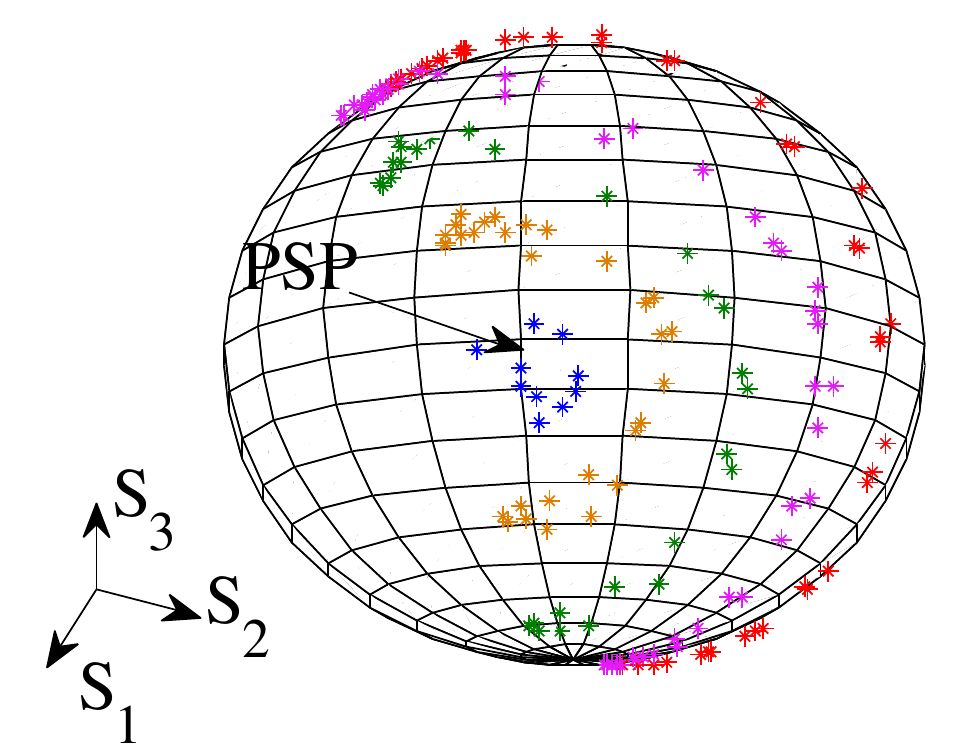}}
\caption{\label{sphereSM}Poincare sphere representation of experimentally measured state of polarization at the output of a single mode fiber amplifier while modulating the pump power. Different colors denote different input states of polarization. The three axes, $S_1$, $S_2$ and $S_3$ are the three Stokes parameters.}
\end{figure}

First, we used a single mode fiber amplifier and measured the SOP at the output as we modulated the pump power at $0.05 Hz$ with modulation amplitude of $360mW$. These pump modulation induces modulation of the temperature of the fiber. The experimental results of the SOP at the output for different input SOP are presented on a normalized Poincare sphere in Fig.~\ref{sphereSM}. As evident, there is one output SOP at the center that is essentially invariant to variations in the pump power, while the other SOP are denoted by increasing concentric circles. The behavior of the increasing concentric SOP in single mode fiber amplifier when varying the temperature is identical to that in passive single mode fibers when varying the wavelength of the input light~\cite{dynamicPSP, dynamicPSP2, dynamicPSP3}. This can be elucidated by resorting to the following simple argument. In general, when the polarization of the light different from the PSP it can be separated into the two PSP components each with a different $k$ vectors. Each component will gain a different phase while propagating along the fiber. The phase difference between them is
\begin{equation}
\Delta \varphi = \frac{ \omega}{c}\int_0^l \Delta n dl,
\label{Eq1}
\end{equation}
where $\omega$ is the light frequency, $l$ is the length of the fiber, $c$ is the speed of light and $\Delta n$ is the difference in the index of refraction for the two components. According to Eq.~\ref{Eq1}, the phase difference $\Delta \varphi$ is linearly proportional to $\omega$ and thereby the wavelength and also to $\Delta n$ and thereby the temperature~\cite{Temp1, Temp2}. So changes in either the wavelength or the pump power which in turn changes the temperature should lead to the same behavior of the SOP~\cite{Kogelnik}. In Poincare sphere representation this yields increasing concentric circles around the PSP for different input SOP when changing the wavelength or the temperature as shown in Fig.~\ref{sphereSM}. The results of Fig.~\ref{sphereSM} clearly indicate that for stable operation, where the output SOP remains essentially constant over a range of temperatures or wavelengths, it is best to choose the PSP as the input with single mode fiber amplifiers.

Next, we replaced the single mode fiber amplifier with a multimode $10m$ long fiber amplifier having core diameter of $20 \mu m$ and numerical aperture of $0.07$ result in a V parameter of $4.1$. In the weekly guiding approximation, such a fiber supports mainly the $LP_{11}$ modes which includes the radial and azimuthal polarizations~\cite{MyRadial, MyRadialAmp}. The detected intensity distribution of the spontaneous emission from the fiber has a doughnut shape, so the Gaussian $HE_{11}$ mode is less amplified. In order to verify that the coupling into the fiber does not change the mode or the polarization, we first performed a control experiment with a short fiber ($7cm$ length) and ensured that the mode and the polarization at the output are the same as those at the input. Then, we injected a linearly polarized light with a uniform intensity distribution into the $10m$ long fiber amplifier, and measured the average SOP across the output beam while modulating the pump power. In order to excite different input polarizations we rotated the SOP of the input beam and in order to excite different input modes we alter the incident angle of the input beam.

\begin{figure}[htb]
\centerline{\includegraphics[width=8.3cm]{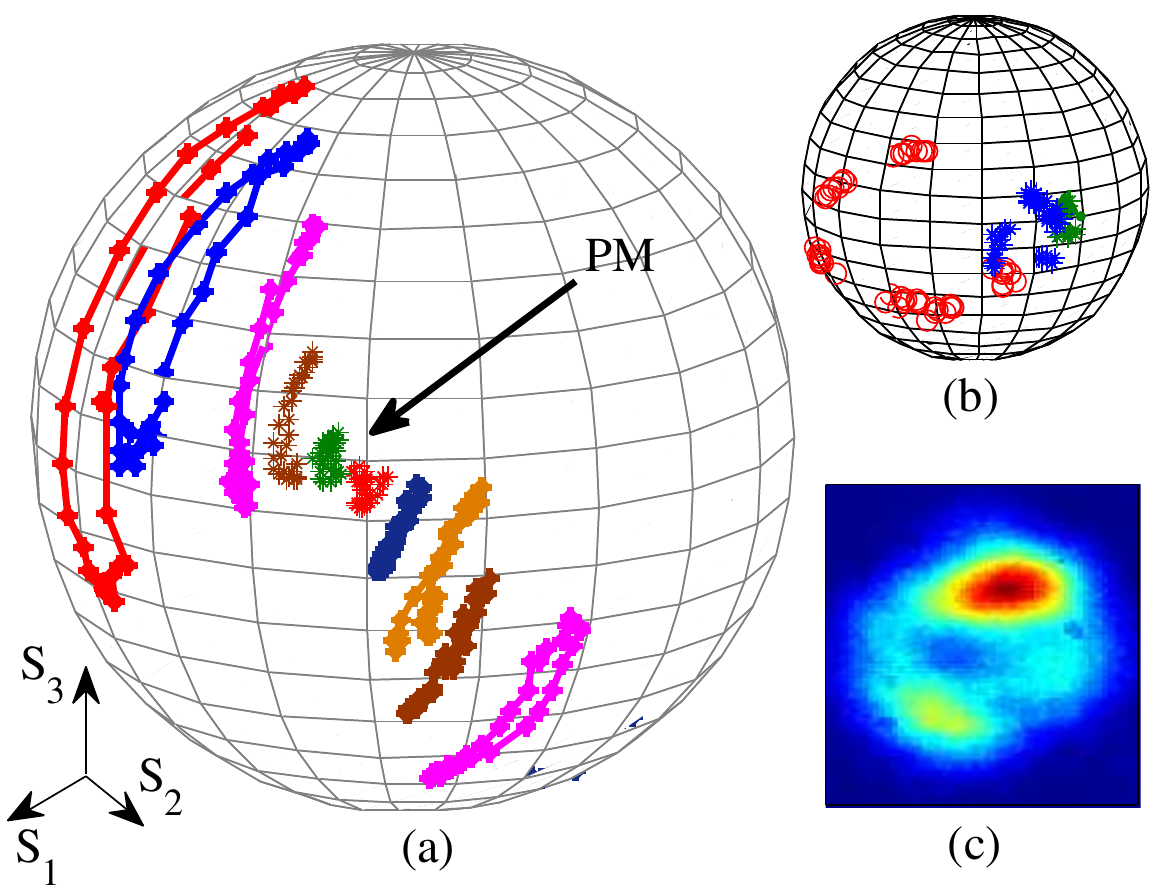}}
\caption{\label{BothEx2}Poincare sphere representation of experimentally measured average state of polarization at the output of a multimode fiber amplifier. (a) while modulating the pump power; (b) while varying the input wavelength. Different colors denote different input states of polarization. The three axes, $S_1$, $S_2$ and $S_3$ are the three Stokes parameters. (c) Present the intensity distribution at the output for the PM.}
\end{figure}

The results are presented in Fig.~\ref{BothEx2}(a). As evident, there is output SOP that is not affected by variations of the pump power, i.e. temperature. Moreover, using a tunable laser, we measured the output SOP as a function of the input wavelength. The results are presented in Fig.~\ref{BothEx2}(b). We found that the SOP that is not affected by variations of the pump power is also not affected by variation of the input wavelength. In multimode fibers this is referred to as the principal mode (PM) of the fiber and the intensity distribution of this mode is presented in Fig.~\ref{BothEx2}(c). Unlike the increasing concentric circles around the PSP location in single mode fiber, here we obtained increasing ellipses whose centers continuously deviate from the PM location as the difference between the input and the PM increases~\cite{}. These results clearly indicate that for stable operation, where the output SOP remains essentially constant over a range of temperatures, it is best to choose the PM as the input with multimode fiber amplifiers.

\begin{figure}[htb]
\centerline{\includegraphics[width=8.3cm]{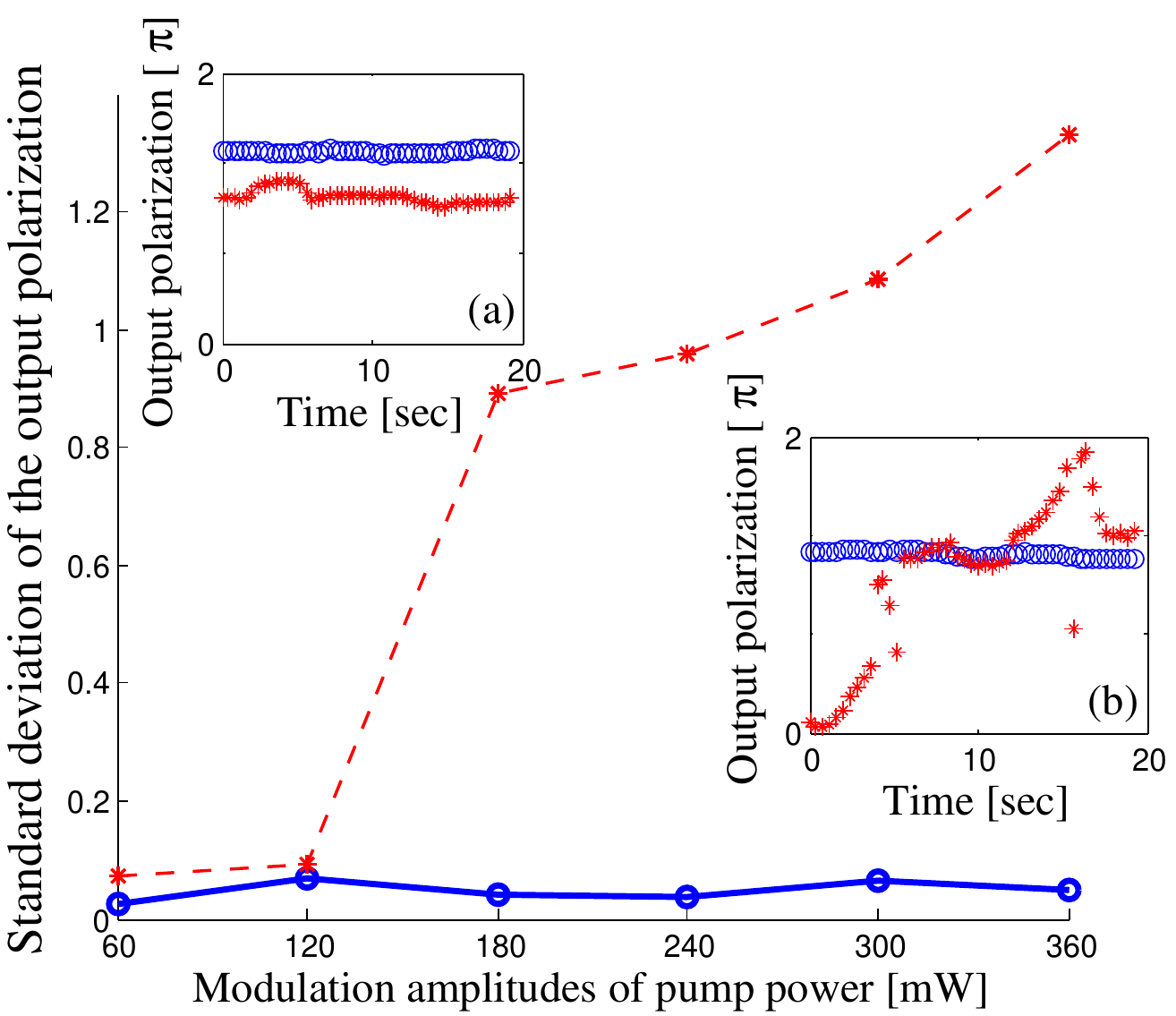}}
\caption{\label{stdofPSP} Experimental results of the standard deviations of the output polarization as a function of the modulation amplitudes of the pump power. Circles (blue) - near the PM; asterisks (red) - far from the PM. Inset (a) shows the polarizations at the output as a function of time when the modulation amplitude of the pump power is $120mW$.  Inset (b) shows the polarizations at the output as a function of time when the modulation amplitude of the pump power is $360mW$.}
\end{figure}

We also performed more detailed measurements near the PM and very far from it in order to quantify the variations of the output SOP as a function of modulation amplitudes of the pump power. Specifically, we measured the output polarization as a function of time at different modulation amplitudes of the pump powers, ranging from $60mW$ to $360mW$. Then, we determined the standard deviations of the output polarizations. The results are presented in Fig.~\ref{stdofPSP}. The solid (blue) curve denotes the standard deviations of output SOP near the PM and the dashed (red) curve denotes those far from it. As evident, when the modulation amplitudes are smaller than $120mW$, the output SOP near and far from the PM are invariant to the amplitude modulations, but above $180mW$ only that near the PM remains invariant. Inset (a) represents the output SOP as a function of time for a modulation amplitude of $120mW$, where both cases of near the PM and far from the PM are insensitive to the pump power; inset (b) represents the output SOP as a function of time for modulation amplitude of $360mW$, where near the PM the SOP at the output remains insensitive to the pump power while far from PM it has significant variations.

To conclude, we investigated the dynamics of the state of polarizations (SOP) at the output from single mode and multimode fiber amplifiers as a function of the input SOP and pump powers. The results revealed that the SOP at the output can vary over a large range as the pump powers (or equivalently the input wavelength or the temperatures of the fiber amplifiers) change. Fortunately, as we clearly demonstrated, such variations can be significantly reduced when the input polarization is at the principal state of polarization (PSP) in single mode fiber amplifiers and principal mode (PM) in multimode fiber amplifiers. Accordingly, the output from the fiber amplifier is more stable and robust. This method can be adopted in many applications where the temperature or the wavelength fluctuate rapidly while the PM or the PSP vary slowly.

%\pagebreak

\end{document}